\documentclass[11pt]{article}
\usepackage[T2A]{fontenc}
\usepackage{ alphalph,etoolbox }
\usepackage{amsthm, amsmath, amssymb, amsbsy, amscd, amsfonts, latexsym, euscript,epsf, bm}

\usepackage{bbold, bm}
\usepackage{epic,eepic}
\usepackage{texdraw}
\usepackage[dvips]{color}
\usepackage{color}
\usepackage{ dsfont }
\usepackage{graphicx}
\usepackage{exscale,relsize}
\usepackage{authblk}
\usepackage{url}
\usepackage{enumerate}
\usepackage{graphicx}

\title{The toy-model of the Boltzmann type equation
for the contact force distribution in disordered packings of particles}
\date{}

\author[1]{Dmitry Grinev\thanks{d.grinyev@uniyar.ac.ru}}
\affil[1]{Centre of Integrable Systems, P.G. Demidov Yaroslavl State University, Yaroslavl, Russia}

\patchcmd{\subequations}{\alph{equation}}{\alphalph{\value{equation}}}{}{}

\begin{document}

\maketitle

\begin{abstract}
The packing of hard-core particles in contact with their neighbors is considered as the simplest model of disordered particulate media. We formulate the statically determinate problem which allows 
analytical investigation of the statistical distribution of contact force magnitude. The toy-model of the Boltzmann type equation for the probability of contact force
distribution is formulated and studied. An experimentally observed exponential distribution is derived.
\end{abstract}

\bigskip

\section{Introduction}\label{sec1}

Understanding rheological properties of packed particles at various spatial scales \cite{bib1} requires application of the apparatus of statistical mechanics. However conventional statistical physics is inadequate when one needs to describe mechanical behavior of  disordered packings of hard-core particles that can be static or driven by external forces \cite{bib2}. There have been various attempts to develop the statistical-mechanical approach for such systems, e.g. see the recent review article \cite{bib3}. Despite these advances there still exists a certain degree of skepticism regarding the possibility of discovering new physical laws that govern mechanical behavior of particulate media. This can be linked to the structural complexity of such materials at different scales. However experimental works \cite{bib4} and computer simulations  \cite{bib5} indicate the existence of phenomena that can and should be treated as physics problems.  In order to to discover physics laws one should consider well-posed problems so that laws can be expected. Our paper does not attempt the ambition of studying the percolation geometry and mechanics of contact forces network \cite{bib6}. It offers a really simple analytical model that produces a Boltzmann type equation for the contact force distribution. This equation can be solved and an experimentally observed exponential distribution is obtained. Possible approaches to develop this simple model are briefly discussed.

\section{The statically determinate problem}\label{sec2}
Let us consider a static array of hard-core particles in contact with their neighbors. The packing is assumed to be an assembly of discrete rigid particles whose interactions with their neighbors are localized at point-like contacts. Therefore the description of the network of interparticle contacts is essential for the understanding of force transmission. We assume that the set of contact points $C_{i}^{\alpha \beta}$
provides the complete geometrical specification for such static packing. We define the centroid of contacts of particle $\alpha$ as
\begin{equation}
R_{i}^{\alpha}=\frac{\sum_{\beta} C_{i}^{\alpha \beta}}{z^{\alpha}}
\end{equation}
where $i = 1, . . . d$ is the Cartesian index,  $z_{\alpha}$ is the coordination number of particle $\alpha$.  The distance between particles $\alpha$ and $\beta$ is defined as the distance
between their centroids of contacts
\begin{equation}
R_{i}^{\alpha \beta}=R_{i}^{\beta}-R_{i}^{\alpha}=r_{i}^{\alpha \beta}-r_{i}^{\beta \alpha}
\end{equation}
where $r_{i}^{\alpha \beta}$ is the $i$-th component of the vector joining the centroid of contact with the contact point i.e.
\begin{equation}
\sum_{\beta} r_{i}^{\alpha \beta}=0
\end{equation}
In $d$ dimensions Newton’s laws of force and couple balance for each particle give us the system of $\frac{Nd(d+1)}{2}$ equations for the interparticle forces $f_{i}^{\alpha \beta}$
\begin{equation}
\sum_{\beta} f_{i}^{\alpha \beta}+g_{i}^{\alpha}=0
\end{equation}
\begin{equation}
f_{i}^{\alpha \beta}+f_{i}^{\beta \alpha}=0
\end{equation}
\begin{equation}
\sum_{\beta} \epsilon_{i k l} f_{k}^{\alpha \beta} r_{l}^{\alpha \beta}+c_{i}^{\alpha}=0
\end{equation}
where $g_{i}^{\alpha}$ is the external body force acting on grain $\alpha$ and $c_{i}^{\alpha}$
is the external body couple which we take to be zero. 
The counting of the number of equations and the number of unknowns allows to formulate the simplest statically determinate problem of force transmission in a static packing. Particles are considered to be perfectly hard, perfectly rough and each particle $\alpha$ has a coordination number $z_{\alpha}= d + 1$ \cite{bib7}. 
Theory which confirms this observation has been proposed for periodic arrays of particles with perfect and zero friction \cite{bib8}. 
What is the statistical distribution of contact forces in a packing of particles? Experimental \cite{bib4} and computer simulation \cite{bib5} studies have demonstrated that the probability of normal contact force acting on a particle contact behaves as 

\begin{equation}
P(f) \propto\left\{\begin{array}{l}
\left(\frac{f}{\langle f\rangle}\right)^{\gamma} \quad \text { if } f<\langle f\rangle \\
e^{\delta\left(1-\frac{f}{\langle f\rangle}\right)} \text { if } f>\langle f\rangle
\end{array}\right.
\end{equation}
where $\langle f\rangle$ is an average contact force, $\gamma$ and $\delta$ are constants.  The aim of this paper is to derive the statistical distribution of contact forces from the first principles. We attempt such derivation by constructing a Boltzmann type equation, which can be solved if the packing of particles  is assumed to be a statically determined i.e. each particle $\alpha$ has a coordination number $z_{\alpha}= d + 1$. Some authors describe such system state as marginal \cite{bib9}.
\section{An integral equation model}\label{sec3}
When a static packing of incompressible particles in contact is subjected to external forces at its boundaries, these forces are transmitted through the contact network. This network is determined by the set of contact points in our model. In order to develop a tractable theory we assume our particles to be mono-disperse spheres in multiple contact greater than or equal to three in 2-D or four in 3-D. Despite this simplification the proposed theory is worthwhile because it offers an analytical solution.
\subsection{The 2-D model}\label{subsec2}
Let us consider a packing in two dimensions where forces $f_{1}, f_{2}$ impinge on a particle which then exerts force $f$ on its neighbor. The average position of the forces is symmetric. 
Let $f$ be in the $x$ direction and use the symbols $f_{1}, f_{2}$ for the $x$ components of the vector forces. Forces can only push and not pull, and the simplest representation of the problem  is as follows

\begin{equation}
P(f)=\int_{0}^{\infty} \mathrm{d} \vec{f}_{1} \int_{0}^{\infty} \mathrm{d} \vec{f}_{2} \delta\left(\vec{f}-\vec{f}_{1}-\vec{f}_{2}\right) 
 P\left(\vec{f}_{1}\right) P\left(\vec{f}_{2}\right)
\end{equation}
After projecting the contact force vectors we have
\begin{equation}
P(f)=\int_{0}^{\infty} \mathrm{d} f_{1} \int_{0}^{\infty} \mathrm{d} f_{2} \int_{0}^{1} \mathrm{~d} \mu \int_{0}^{1} \mathrm{~d} \lambda 
\delta\left(f-\lambda f_{1}-\mu f_{2}\right) P\left(f_{1}\right) P\left(f_{2}\right)
\end{equation}
which can be transformed into
\begin{equation}
P(f)= \int_{0}^{\infty} \mathrm{d} f_{1} \int_{0}^{\infty} \mathrm{d} f_{2} \int_{0}^{1} \mathrm{~d} \mu 
 \int_{0}^{1} \mathrm{~d} \lambda P\left(\lambda f_{1}\right) P\left(\mu f_{2}\right)
\end{equation}
This equation has the solution in the following form
\begin{equation}
P(f)=\frac{f}{p^{2}} e^{-\frac{f}{p}}
\end{equation}
which has been normalised and where $p = \frac{\bar{f}}{2}$ $\bar{f}$ corresponds to the mean force. The distribution $P(f)$ is exponential for large values of $f$ and goes to zero at small $f$.

\subsection{The 3-D model}\label{subsec2}

In 3-D the coordination number of particles is greater than or equal to four. Using the same framework of simplifications, the three-dimensional model can be written as

\begin{equation}
f=\frac{1}{3} f_{1}+\frac{1}{3} f_{2}+\frac{1}{3} f_{3}
\end{equation}
After applying the Fourier-transform
\begin{equation}
 P(f) = \frac{1}{(2\pi)^3} \int d^3k \mathcal{P}(k) e^{ikf}
\end{equation}
 we obtain
\begin{equation}
\mathcal{P}(k)=\mathcal{P}^{3}\left(\frac{k}{3}\right)
\end{equation}
when again
\begin{equation}
\mathcal{P}(k)=e^{i \frac{k}{p}}
\end{equation}
After applying the inverse Fourier-transform 
\begin{equation}
 \mathcal{P}(k)= \int d^3f P(f) e^{-ikf}
\end{equation}
we obtain
\begin{equation}
P(f)=\delta(f-p)
\end{equation}
The blurring process is always employed in the form of a ``toy model''. In this case the angular effects can be represented by three ‘‘direction cosines’’  $\lambda_{i}$ so the force balance equation in the form 
\begin{equation}
f=\lambda_{1}^{2} f_{1}+\lambda_{2}^{2} f_{2}+\lambda_{3}^{2} f_{3}
\end{equation}
offers an analytic solution 
\begin{equation}
P(f)=\left(\int_{-\infty}^{\infty} \int_{0}^{1} \mathcal{P}\left(\lambda^{2} k\right) \mathrm{d} k \mathrm{~d} \lambda\right)^{3}
\end{equation}
Using $\lambda^{2}k=\mu$ we obtain
\begin{equation}
P(f)=\left(\int_{0}^{K} \mathcal{P}(\mu) \frac{\mathrm{d} \mu}{2 \mu^{\frac{1}{2}} k^{\frac{1}{2}}}\right)^{3}
\end{equation}
Applying the Fourier-transform to this expression we obtain
\begin{equation}
\mathcal{P}(k)=\frac{4 p^{\frac{3}{2}}}{(k-i p)^{\frac{3}{2}}}
\end{equation}
so we have
\begin{equation}
P(f)=\int \frac{4 p^{\frac{3}{2}} e^{i k f}}{(k-i p)^{\frac{3}{2}}} \mathrm{~d} k=4 p^{\frac{3}{2}} e^{-\frac{f}{p}} \int \frac{e^{i J f}}{J^{\frac{3}{2}}} \mathrm{~d} J
\end{equation}
Integration of this expression gives the normalised distribution
\begin{equation}
P(f)=\frac{\sqrt{\pi}}{2} f^{\frac{1}{2}} p^{\frac{3}{2}} e^{-\frac{f}{p}}
\end{equation}
where $p\propto \bar{f}$ and the proportionality constant depends on the exponent of the power law rise at low forces. 
Indeed, a number of improvements must to be done in order to derive this expression with experimentally observed coefficients. However, as the starting point, the use of this simple model appears to be justified. 
\section{Discussion}\label{sec3}
We proposed the simplest Boltzmann type equation for the probability distribution of contact forces in two and three dimensions. This equation can be solved under some approximations and an experimentally observed 
exponential distribution of contact forces can be obtained. Other proposed approaches\cite{bib10}-\cite{bib12} employ entropy maximization or functional minimization concepts. These models produce elements of the empirically observed probability distribution function. However they are not derived from first-principles but are developed by analogy with other entropic systems. We hope that our approach of studying
the Boltzmann type equation for the probability distribution of contact forces can serve as the foundation for future research. In particular one can develop it further to account for the presence of structural disorder at various spatial scales and obtain the statistics of so called ``force chains'' which are observed in experiment\cite{bib4}. 

\section{Conflict of Interest} 
The author declares that he has no conflict of interest.
\section{Acknowledgments}

The work on sections $3$ and $4$ was supported by the Russian Science Foundation (grant No. $21-71- 30011$). The work on sections $1$ and $2$ was carried out within the framework of a development programme for the Regional Scientific and Educational Mathematical Center of the Yaroslavl State University with financial support from the Ministry of Science and Higher Education of the Russian Federation (Agreement on provision of subsidy from the federal budget No. $075-02-2022-886$). The author thanks Prof. R. C. Ball for many stimulating discussions. Last but not least, valuable comments and suggestions by an anonymous Referee that helped improve the clarity are gratefully acknowledged.

\end{document}